\documentclass[12pt]{iopart}
\usepackage{graphicx}
\usepackage{graphics}
\usepackage{dcolumn}
\usepackage{bm}
\usepackage{epsfig}

\begin{document}

\title[Identified particle ratios from STAR in Au+Au collisions @ $\sqrt{s_{NN}}$ = 200 GeV]{Identified particles at large transverse momenta in STAR in Au+Au collisions @ $\sqrt{s_{NN}}$ = 200 GeV}

\author{M A C Lamont\dag for the STAR Collaboration\footnote[3]{For the full author list and acknowledgements, see Appendix ``Collaborations" in this volume.}}

\address{\dag\ School of Physics and Astronomy, The University of Birmingham, Edgbaston, Birmingham B15 2TT, UK}

\begin{abstract}

We report measurements of the ratios of identified hadrons ($\pi$,K,p,$\Lambda$) in Au+Au collisions at $\sqrt{s_{NN}}$ = 200 GeV as a function of both collision centrality and transverse momentum ($p_T$).  Ratios of anti-baryon to baryon yields are independent of $p_T$ within 2$<$$p_T$$<$6 GeV/c indicating that, for such a range, our measurements are inconsistent with theoretical pQCD calculations predicting a decrease due to a stronger contribution from valence quark scattering.  For both strange and non-strange species, a strong baryon enhancement relative to meson yields is observed as a function of collision centrality in this intermediate $p_T$ region, leading to $p/\pi$ and $\Lambda$/K ratios greater than unity.  The nuclear modification factor, $R_{cp}$ (central relative to peripheral collisions), is used to illustrate the interplay between jet quenching and hadron production.  The physics implications of these measurements are discussed with reference to different theoretical models~\cite{Fries, Vitev, Greco}\footnote[2]{For discussion on strange particle correlations which were presented at this conference, refer to~\cite{Ying}.}.

\end{abstract}



\vspace{-0.6 cm}

\section{Introduction}
At large transverse momenta, particle production in ultra-relativistic heavy ion collisions is dominated by jet fragmentation, and at low momenta by soft particle production which can be well described by hydrodynamical models~\cite{Hydro}.  In the intermediate $p_T$ region, (2 $<$ $p_T$ $< \sim$ 6 GeV/c), the mechanisms of particle production are still to be investigated.  It has been reported that the $\overline{p}/p$ ratio is approximately constant as a function of $p_T$ up to 4 GeV/c~\cite{PHENIX_pbarp}, in disagreement with pQCD calculations which predict a decrease with increasing $p_T$~\cite{Vitev}.  Also, the $p/\pi$ ratio rises with increasing $p_T$ up to 4 GeV/c, where it reaches unity~\cite{PHENIX_pbarp}.  This is much larger than the ratio found in elementary collisions~\cite{PbarP_elem}.  It has also been shown that at higher $p_T$, particle production does not scale with the number of binary collisions in the most central Au+Au collisions, but rather is suppressed~\cite{STAR_jetquenching, PHENIX_jetquenching}.  The particle dependence of this quantity, $R_{cp}$ (the ratio between the yields in central and peripheral collisions as a function of $p_T$), exhibits a difference between the lighter ($\pi,K$) and the heavier particles ($\Lambda, p$).

In this paper we present the $\overline{p}/p$ and $\overline{\Lambda}/\Lambda$ ratios as a function of $p_T$ in the most central collisions, extending the $p_T$ coverage to 6 GeV/c.  The $\overline{p}/\pi^{-}$ and $\Lambda/K^0_s$ ratios are presented as a function of both $p_T$ and the centrality of the collision, extending the measurement of the baryon/meson ratio reported in \cite{PHENIX_pbarp}.  The $R_{cp}$ ratios for the $\Xi$ and $K^*(892)$ are also presented, giving further information on the baryon and meson production differences.

\section{Analysis and Results}

In this analysis, 1.6 x 10$^6$ minimum-bias triggers and 1.5 x 10$^6$ central triggers in the STAR detector system were used~\cite{STAR_Trigger}.  The charged hadrons ($p, \pi$) were identified using a small acceptance ($\Delta\eta < $0.3, $\Delta\phi $ = 20$^\circ$) CsI based Ring Imaging CHerenkov detector (RICH), located at mid-rapidity in the STAR setup~\cite{STAR_RICH}, extending the $p_T$ coverage of the particle identification of the $\pi$ and $K$ up to 3 GeV/c, and the $p$ to 5 GeV/c, well beyond the range achieved through specific ionisation in the Time Projection Chamber (TPC)~\cite{STAR_TPC}.  Meanwhile, the strange hadrons ($\Lambda$ and $K^{0}_{s}$) are identified at mid-rapidity ($|y| <$ 1.0) in the TPC through their charged decay channels : $\Lambda \rightarrow p+\pi^{-} (64\%)$, $\overline\Lambda \rightarrow \overline{p}+\pi^{+} (64\%)$ and $K^{0}_{s} \rightarrow \pi^+ + \pi^- (69\%)$~\cite{STAR_V0}.

\begin{center}
\begin{figure}[h!]
\vspace{-0.70 cm}
\begin{minipage}[h!]{0.48\textwidth}
\hspace{0.0 cm}
 \vspace{0.6 cm}
 \includegraphics[width=1.0\textwidth,height=0.66\textwidth]{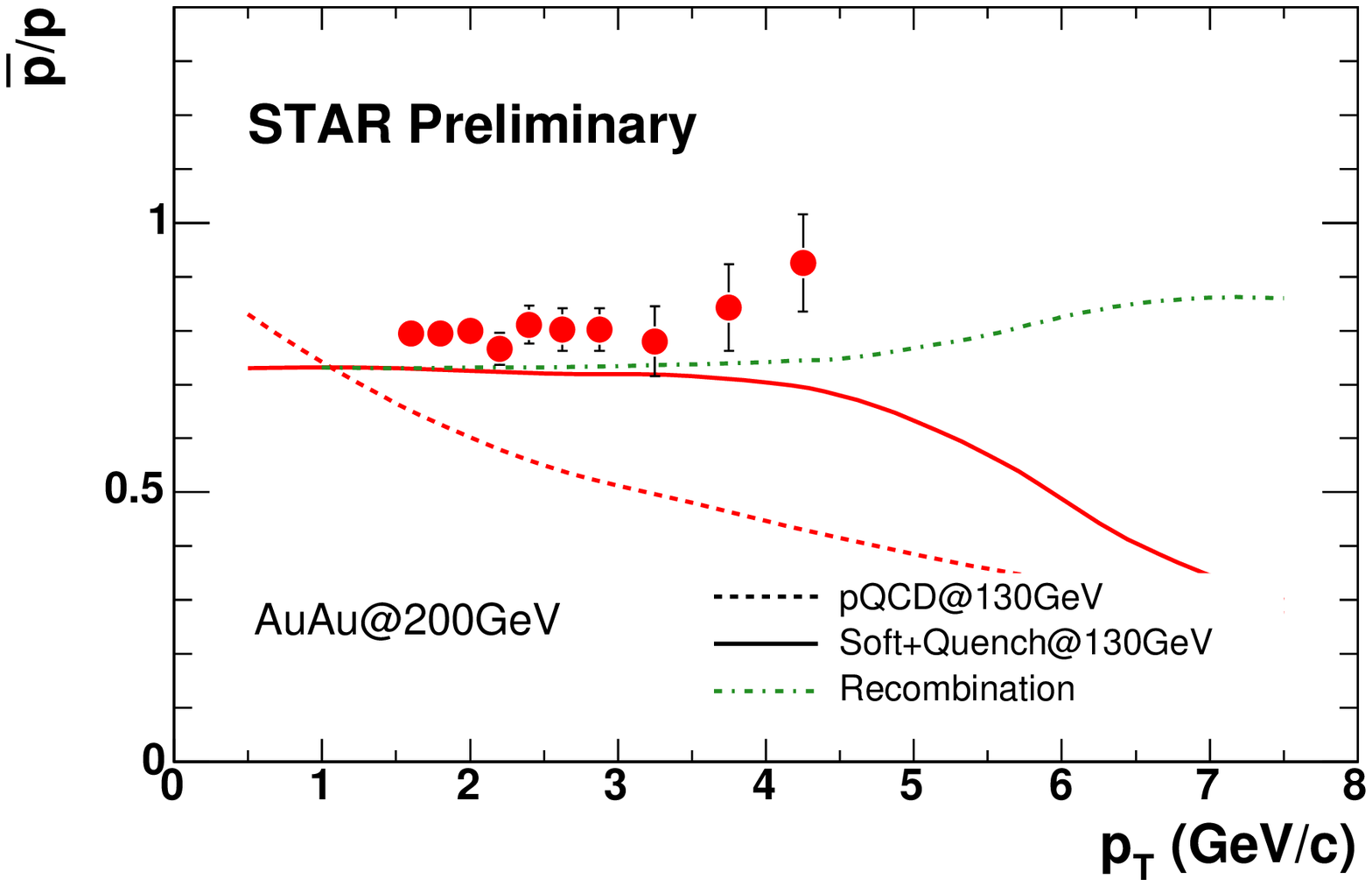} 
 \vspace{0.2cm}
\end{minipage}\hfill
\begin{minipage}[h!]{0.48\textwidth}
 \hspace{-1.5 cm}
 \vspace{0.2 cm}
 \includegraphics[width=1.0\textwidth,height=0.66\textwidth]{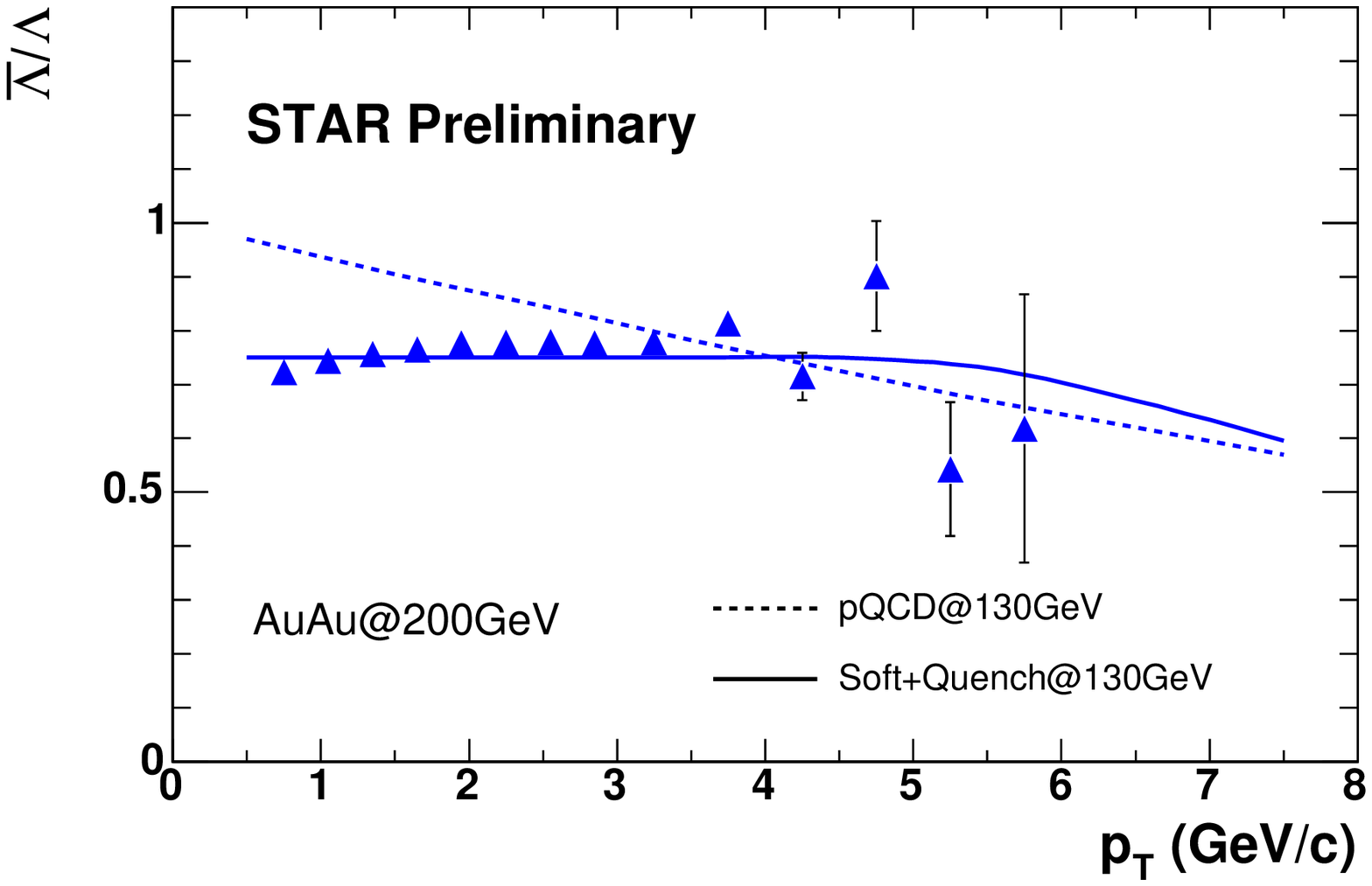} 
 \vspace{1.15 cm}
\end{minipage}
 \vspace{-1.6 cm}
 \hspace{-0.8cm}
\caption{The $\overline{p}/p$ and $\overline{\Lambda}/\Lambda$ ratios vs $p_T$, together with different theoretical models.}
\label{Fig:BbarB}
\end{figure}
\end{center}

The $\overline{p}/p$ and $\overline{\Lambda}/\Lambda$ ratios are presented as a function of $p_T$ for the most central collisions (0-10$\%$) in Figure \ref{Fig:BbarB} \footnote[3]{The $\overline{p}/p$ ratio presented here is different to that presented at the previous conference in this series~\cite{PbarP_Kunde}, where the ratio showed a clear decrease with increasing $p_T$.  Since then, experimental effects have been better modelled (mainly space charge distortions) leading to an improved analysis.}.  Along with the data, predictions are also plotted for the $\overline{p}/p$ ratio from the `Soft+Quench' (solid) model (and pQCD (dashed) model from the same authors) at 130 GeV~\cite{Vitev, Vitev2}, and a recombination (dash-dot) model at 200 GeV~\cite{Fries}.  Predictions from the `Soft+Quench' and pQCD models at 130 GeV are also shown for the $\overline{\Lambda}/\Lambda$ ratio~\cite{Vitev2}.

\vspace{-0.5 cm}

Allowing for the discrepancies in energy between some of the model predictions and the data, both the recombination and `Soft+Quench' models are in agreement with the data, whilst the pQCD prediction shows a decrease in the ratio over all $p_T$.  This pQCD calculation therefore fails to describe the data in this $p_T$ range, though the uncertainties in PDFs and fragmentation functions need to be addressed.

\begin{center}
\begin{figure}[h!]
\vspace{-0.90 cm}
\begin{minipage}[h!]{0.48\textwidth}
\hspace{2.0 cm}
 \vspace{0.6 cm}
 \includegraphics[width=0.85\textwidth,height=0.85\textwidth]{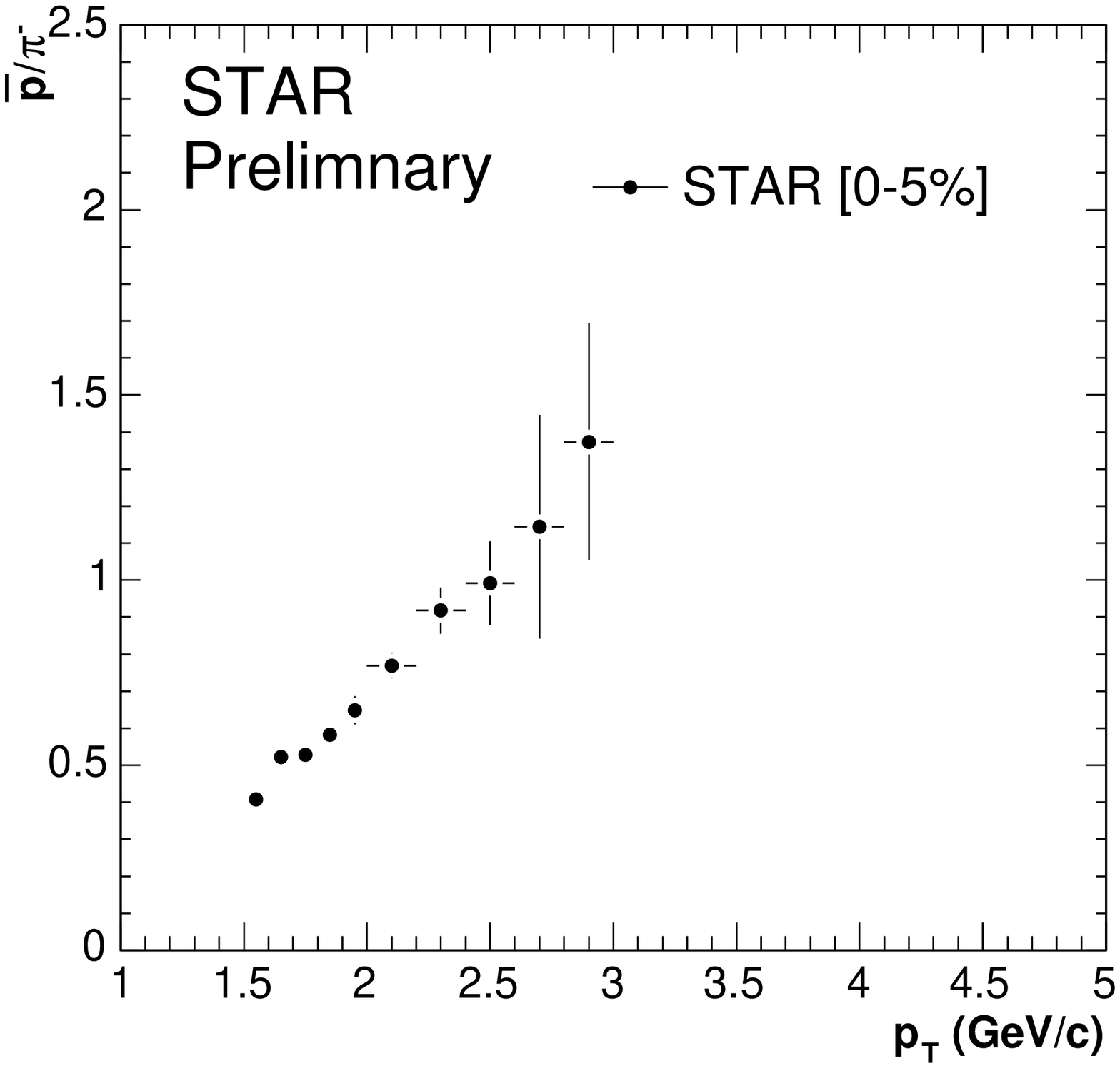} 
 \vspace{0.2cm}
\end{minipage}\hfill
\begin{minipage}[h!]{0.48\textwidth}
 \hspace{-0.6 cm}
 \vspace{0.2 cm}
 \includegraphics[width=0.87\textwidth,height=0.87\textwidth]{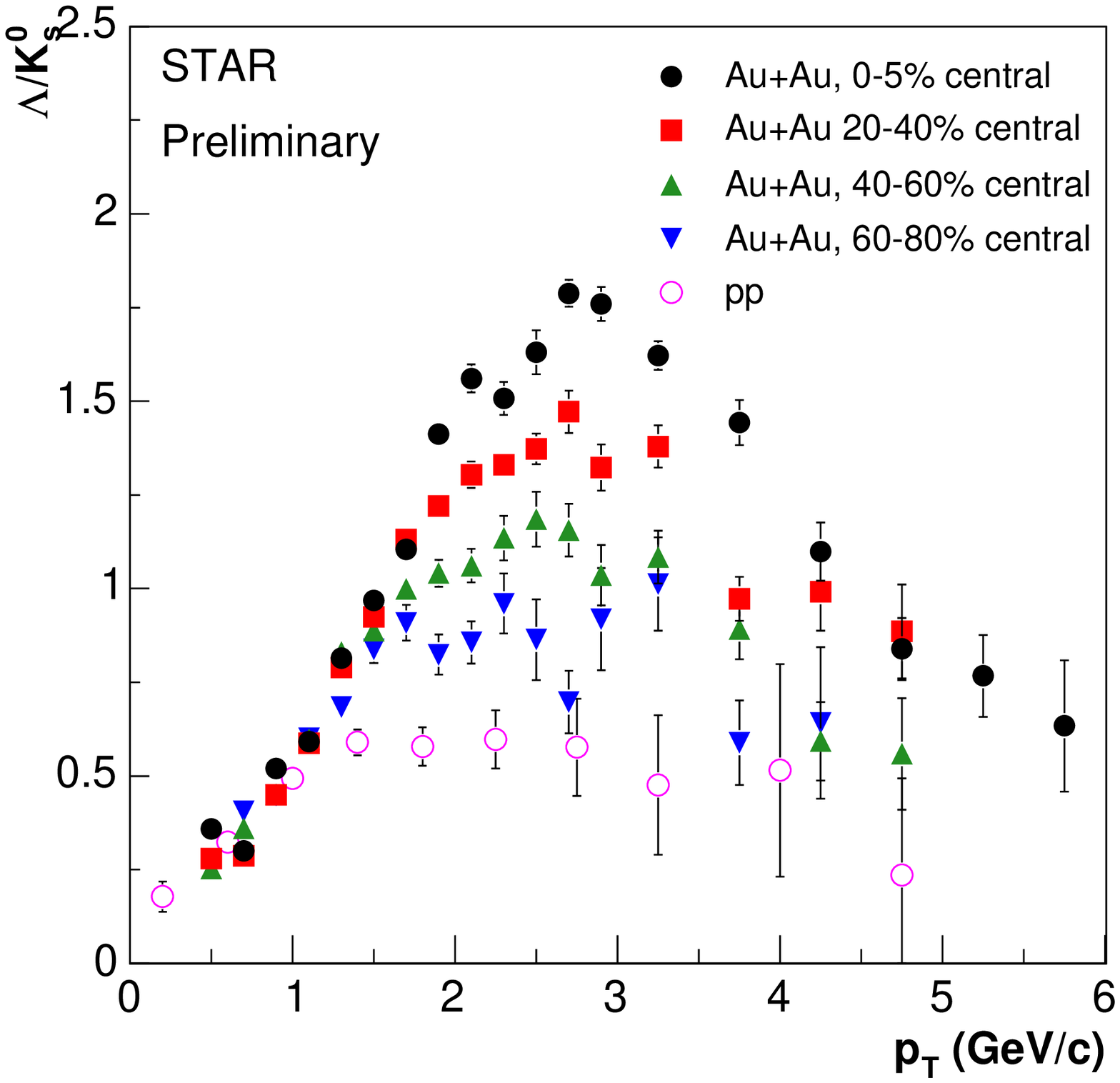} 
 \vspace{1.15 cm}
\end{minipage}
 \vspace{-1.4 cm}
 \hspace{-0.8cm}
\caption{Baryon/meson ratios as a function of $p_T$: On the left, $\overline{p}/\pi^-$ for the 5$\%$ most central collisions; on the right, $\Lambda/K^0_s$ as a function of collision centrality.}
\label{Fig:BaryonToMeson_Data}
\end{figure}
\end{center}

The baryon/meson ratios can also be studied in both the non-strange and strange sector via the $\overline{p}/\pi^-$ and $\Lambda/K^0_s$ ratios and are presented in Figure \ref{Fig:BaryonToMeson_Data}.  For the $\overline{p}/\pi^-$ ratio, which is in agreement with earlier measurements~\cite{PHENIX_pbarp}, the $p_T$ range is limited to 3 GeV/c due to the need for unambiguous identification of the $\pi^-$, whilst the $p_T$ coverage of the $\Lambda/K^0_s$ ratio is governed by (lack of) statistics.  

\vspace{-0.5 cm}

\begin{figure}[h!]
\vspace{-0.5 cm}
\begin{center}
\epsfig{figure=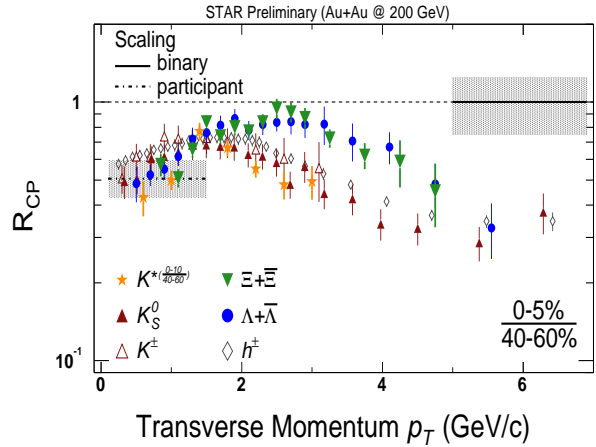, width=0.56\textwidth, height=0.4\textwidth}
\end{center}
\vspace{-0.4 cm}
\caption{The $R_{cp}$ for different mass hadrons.  A clear difference exists between baryons and mesons.}
\label{Fig:Rcp}
\end{figure}

The $R_{cp}$ measurements for the $K^*(892)$ and the $\Xi$ are plotted in Figure \ref{Fig:Rcp}, along with the $\Lambda$ and kaon, where the shaded regions show what is expected for participant and binary scaling respectively.  The variable $R_{cp}$, plots the relative yield of particles in the most central Au+Au collisions to the yield in peripheral Au+Au collisions, normalising each yield to the number of binary collisions, as a function of $p_T$.

\vspace{-0.5 cm}

\section{Discussion}

The $\Lambda/K^0_s$ ratio shows a smooth increase with the number of participating nucleons, from minimum bias p+p collisions (at 200 GeV/c), up to the most central Au+Au collisions.  For all centralities, there appears to be a constant increase of the ratio at low $p_T$, whereupon the ratio in that collision system plateaus and then falls off.  The $\Lambda/K^0_s$ (and hence the baryon/meson) ratio is greater than unity in the most central data for the range 2 $< p_T < $ 6 GeV/c.  The data appears to tend to the same value for all centralities in the range $p_T$ $\sim$ 5-6 GeV/c, but is not yet down to the level exhibited in p+p collisions.  This feature is also present in the $R_{cp}$ ratios in Figure \ref{Fig:Rcp}.  With the addition of the $K^*(892)$ and $\Xi$ data, it is evident that baryons and mesons are suppressed at different transverse momenta, and the differences previously reported between the $\pi$ and $p$~\cite{PHENIX_pbarp} and $K^0_s$ and $\Lambda$~\cite{STAR_RCP} are not due to a mass effect.  We note that the $R_{cp}$ for the $\Lambda$ and $K^0_s$ come together at approximately the same $p_T$ as the $\Lambda/K^0_s$ ratios come together, perhaps indicating the onset of particle production from string fragmentation.

\begin{center}
\begin{figure}[h!]
\vspace{-0.5 cm}
\begin{minipage}[h!]{0.48\textwidth}
\hspace{-0.5 cm}
 \vspace{0.6 cm}
 \includegraphics[width=1.1\textwidth,height=0.55\textwidth]{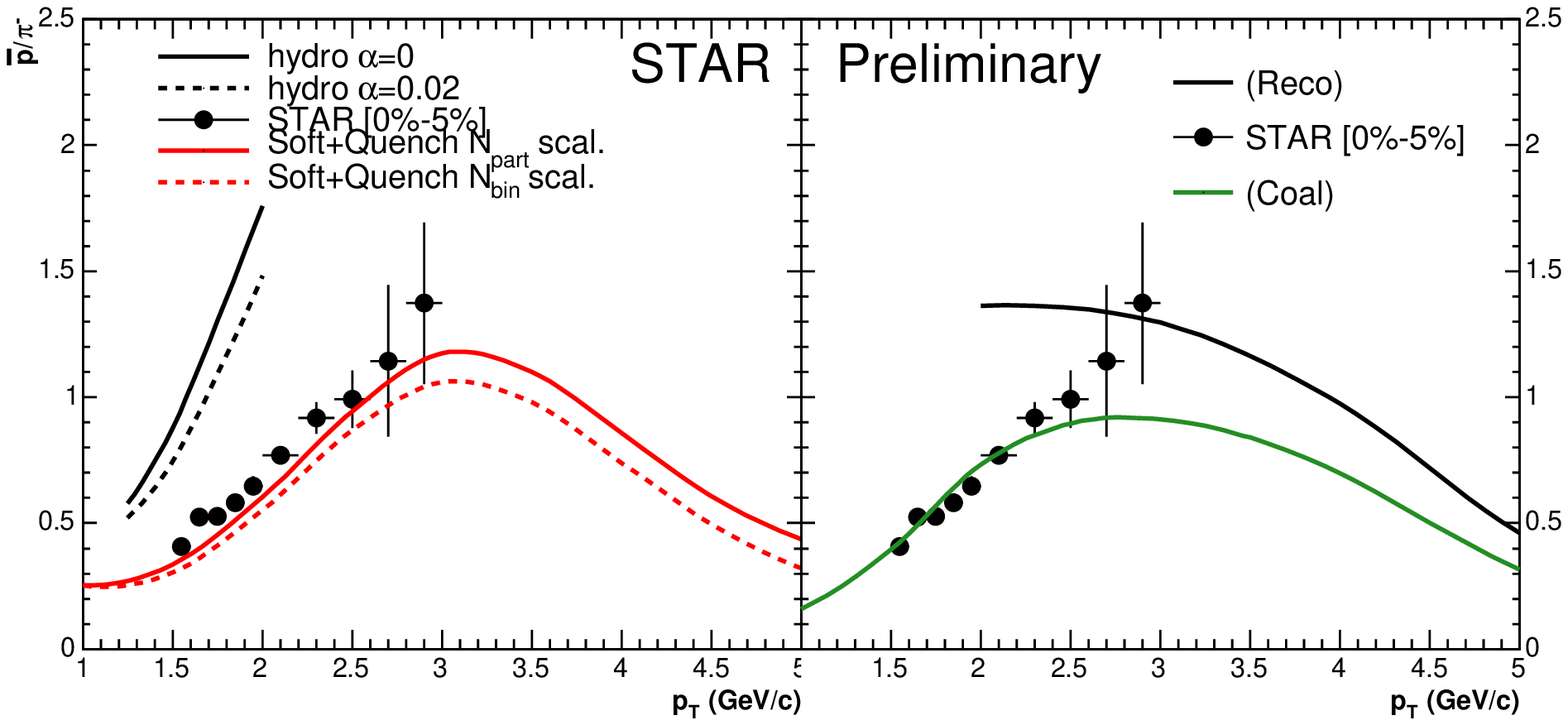} 
\end{minipage}\hfill
\begin{minipage}[h!]{0.48\textwidth}
 \hspace{-0.95 cm}
 \vspace{-0.055 cm}
 \includegraphics[width=1.1\textwidth,height=0.55\textwidth]{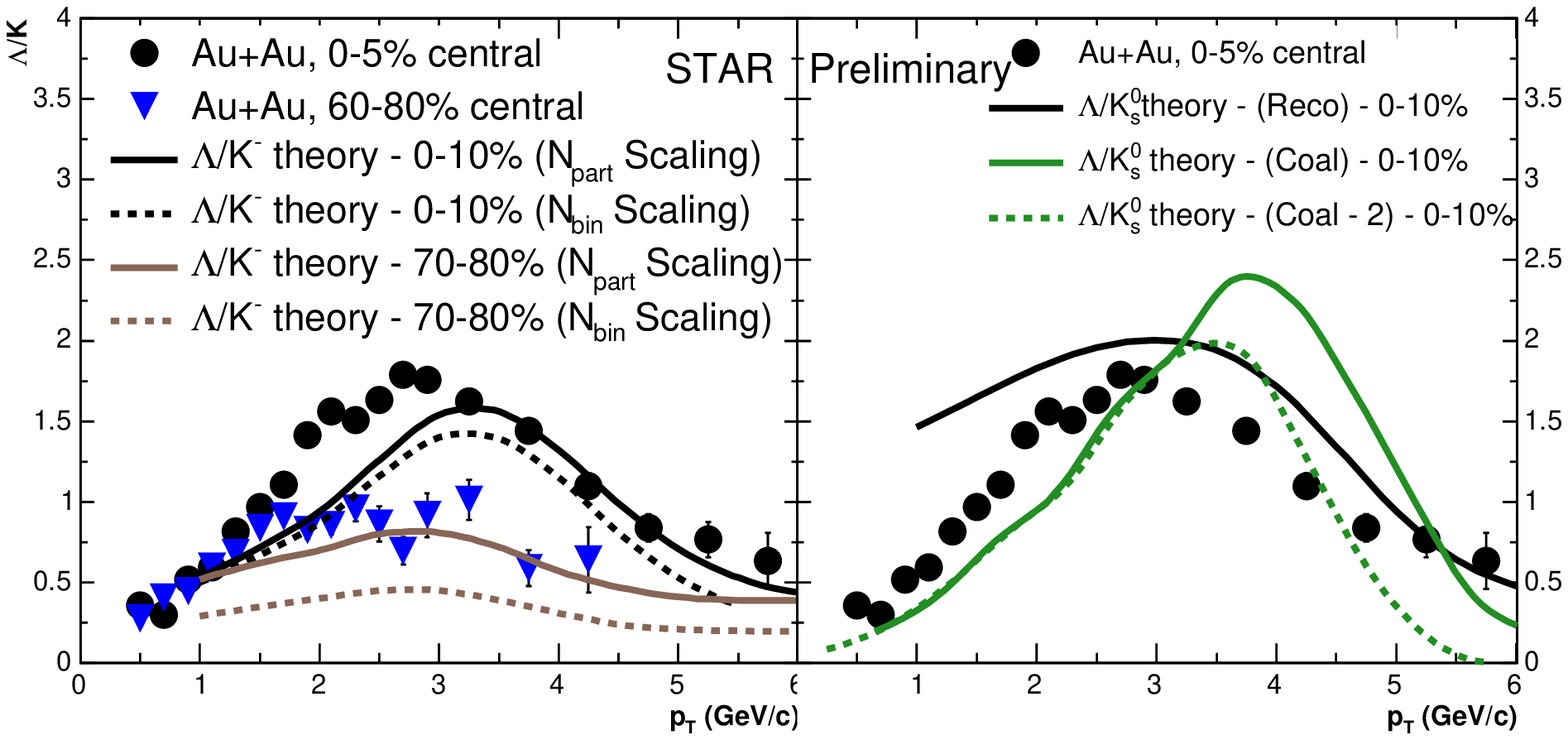} 
 \vspace{0.65cm}
\end{minipage}
 \vspace{-0.7 cm}
 \hspace{-0.5cm}
\caption{Baryon/meson ratios plotted with various theoretical models.  The two leftmost plots are for the $\overline{p}/\pi^-$ ratio and the rightmost two plots for the $\Lambda/K^0_s$ ratio.}
\label{Fig:BaryonToMeson_Models}
\end{figure}
\end{center} 

Theoretical models which use various mechanisms for baryon production such as gluon junctions (Soft+Quench)~\cite{Vitev, Vitev_Private}, recombination~\cite{Fries} and quark coalescence~\cite{Greco} are compared to the data in Figure \ref{Fig:BaryonToMeson_Models}.  The two different coalescence comparisons are for with (Coal) and without (Coal-2) the coalescence mechanism from within the same model.  This is strongly dependent upon the wavefunction used for the $\Lambda$ in the analysis~\cite{Greco_Private}.  The recombination and coalescence models require a large parton density and though they appear to be more natural than the gluon junction mechanism, all the models are in quite good agreement with the central data.  In order to understand the data further and differentiate between the models, we require realistic calculations as a function of centrality.

\section{Summary}

In summary, we have shown that both the $\overline{p}/p$ and $\overline{\Lambda}/\Lambda$ ratios are independent of $p_T$ up to 4.5 and 6 GeV/c respectively.  This finding is inconsistent with the pQCD model presented in the $p_T$ range considered.  We have also found that the $\Lambda/K^0_s$ ratio increases smoothly as a function of increasing centrality in the intermediate $p_T$ region, to a plateau value of approximately 1.8 in the most central collisions.  This ratio then falls again, reducing to below unity at a $p_T$ of approximately 5 GeV/c, though the ratio is still above the value from p+p collisions at the same energy.  We also presented the $R_{cp}$ values for the $K^*$ and the $\Xi$, which showed that the observed suppression does not originate from the hadrons mass, but rather that the suppression is different for baryons and mesons.

\end{document}